\documentclass[preprint,prd,nofootinbib,tightenlines]{revtex4}
\usepackage{graphicx}
\usepackage{bm}
\usepackage{multirow}

\begin{document}
\baselineskip=15pt \parskip=3pt

\vspace*{3em}

\preprint{}

\title{\boldmath Effect on Higgs Boson Decays from Large Light-Heavy \\
Neutrino Mixing in Seesaw Models}
\author{Jyong-Hao Chen}
\affiliation{Department of Physics and Center for Theoretical Sciences, \\
National Taiwan University, Taipei 106, Taiwan}
\author{Xiao-Gang He}
\affiliation{Department of Physics and Center for Theoretical Sciences, \\
National Taiwan University, Taipei 106, Taiwan}
\author{Jusak Tandean}
\affiliation{Department of Physics and Center for Theoretical Sciences, \\
National Taiwan University, Taipei 106, Taiwan}
\author{Lu-Hsing Tsai}
\affiliation{Department of Physics and Center for Theoretical Sciences, \\
National Taiwan University, Taipei 106, Taiwan}

\date{\today $\vphantom{\bigg|_{\bigg|}^|}$}

\begin{abstract}
In seesaw models with more than one generation of light and heavy neutrinos, $\nu$~and~$N$,
respectively, it is possible to have sizable mixing between them for heavy-neutrino masses
of order 100\,GeV or less.
We explore this possibility further, taking into account current experimental constraints,
and study its effect on Higgs-boson decays in the contexts of seesaw models of types~I and~III.
We find that in the type-I case the Higgs decay into a pair of
light and heavy neutrinos, $h\to\nu N$, could increase the total Higgs width in the standard
model by up to almost 30\% for a relatively light Higgs-boson, which would significantly
affect Higgs searches at the LHC.
The subsequent prompt decay of $N$ into three light fermions makes this Higgs decay
effectively a four-body decay.
We further find that, in the presence of the large light-heavy mixing, these four-body Higgs decays
can have rates a few times larger than their standard-model counterparts and therefore could
provide a~potentially important window to reveal the underlying seesaw mechanism.
\end{abstract}


\maketitle

\section{Introduction}

Various experiments have now established that neutrinos have mass and mix with each
other~\cite{pdg}.
The masslessness of the neutrinos in the minimal standard model (SM) implies that one has to go
beyond it to account for this observation.
Among a number of possibilities that have been
proposed~\cite{models,seesaw1,Foot:1988aq,large-mix,Pilaftsis:1991ug,de Gouvea:2007uz},
the most popular are the seesaw scenarios in which new particles are introduced with masses
sufficiently large to make the neutrino masses small.

In the so-called type-I and type-III seesaw models~\cite{seesaw1,Foot:1988aq}, the heavy
particles responsible for giving mass to the light neutrinos are neutral fermions,
often referred to as heavy neutrinos.
Whether the seesaw mechanism can be probed at colliders crucially depends not only on
the masses of the heavy neutrinos (as well as their charged partners in the case of type III),
but also on the strength of their interactions with SM particles, specifically
the mixing between the heavy neutrinos, $N$, and the light ones,~$\nu$.

With only one generation of the neutrinos, the size of this light-heavy mixing is of order
the square root of their mass ratio,~$(m_\nu^{}/m_N^{})^{1/2}$.
Since the light-neutrino mass must be less than an~eV or so, the mixing would be very
small, less than $10^{-5}$ even for $m_N^{}$ of order~100\,GeV.
This would make it challenging to test the seesaw mechanism, especially in the type-I model,
at colliders.
However, in the presence of more than one generation of light and heavy neutrinos, there are
circumstances in which the mixing can be much
larger~\cite{large-mix,Pilaftsis:1991ug,de Gouvea:2007uz,Kersten:2007vk},
offering greater hope of observing its effects on various processes.
The combination of such large mixing, with \,\mbox{$m_N^{}\sim100$\,GeV,}\, and the tiny
light-neutrino masses can occur naturally if the underlying theory has some symmetry that is
slightly broken~\cite{Kersten:2007vk}.

A recent study~\cite{He:2009ua} has explored this possibility of large light-heavy mixing
further and considered specific examples in both seesaw scenarios of types~I and~III.
That study also examined some of the implications of the large light-heavy mixing for
the single production of the heavy leptons at the LHC via channels such as
\,$q\bar q'\to W^*\to l N$\, and found that there are interesting prospects for detecting
these heavy particles at the LHC.

In the present paper, we consider additional processes where it may be possible to probe
the large-mixing effects.
In particular, we apply some of the results obtained in Ref.~\cite{He:2009ua} for types-I
and~-III seesaw to the decays of the Higgs boson into a light ordinary fermion plus one of
the new heavy leptons.
We show that, with the light-heavy mixing as large as allowed by currently available experimental
data, some of these new decay modes of the Higgs boson could give rise to sizable modifications
of its decay branching ratios in the SM and therefore could significantly alter Higgs searches at
the LHC or other colliders.
On~the other hand, the new decay modes could serve to open a window to the underlying seesaw
mechanism if the Higgs boson is discovered and its decay modes are well measured.

\section{Large light-heavy mixing in type-I seesaw}

In the type-I scenario, the seesaw mechanism is realized by introducing right-handed neutrinos
that are singlets under the SM gauge groups and can therefore have large Majorana
masses~\cite{seesaw1}.  Following Ref.~\cite{He:2009ua}, we assume for definiteness that
there are three of these heavy neutrinos, $N_{iR}$, responsible for giving mass to
the three left-handed light neutrinos,~$\nu_{iL}^{}$.
The relevant Lagrangian describing the masses of the neutrinos can be expressed as
\begin{eqnarray}
{\cal L} \,\,=\,\,
- \bar N_{iR}^{} (Y_D^{})_{ij}^{} \tilde H^\dagger L_{jL}^{}
- \mbox{$\frac{1}{2}$} \bar N_{iR}^{} (M_N^{})_{ij}^{} \bigl(N_{jR}^{}\bigr)^c
\,\,+\,\, {\rm H.c.} ~,
\end{eqnarray}
where summation over \,$i,j=1,2,3$\, is implied,  $Y_D^{}$ is the (3$\times$3) Yukawa coupling
matrix, \,$\tilde H=i\tau_2^{}H^*$\,  with $\tau_2^{}$ being the usual second Pauli matrix and
\,$H=\bigl(\phi^+~~(v+h+i\eta)/\sqrt2\bigr){}^{\rm T}$\, the Higgs doublet, $v$~its vacuum
expectation value,  \,$L_{iL}^{}=\bigl(\nu_{iL}^{}~~l^-_{iL}\bigr){}^{\rm T}$\, is
the left-handed lepton doublet,  $M_N$~is the Majorana mass matrix, and $(N_{iR}^{})^c$\,
denotes the charge conjugate of~$N_{iR}$.
The resulting seesaw mass terms are given by
\begin{eqnarray} \label{Lmass}
{\cal L}_{\rm mass}^{} \,\,=\,\, -\mbox{$\frac{1}{2}$}
\Bigl( \overline{\bigl(\nu_L^{}\bigr)^{c}} \hspace{3ex} \bar N_R^{} \Bigr)
M_{\rm seesaw}^{} \left( \begin{array}{c} \nu_L^{} \vspace{1ex} \\
\bigl(N_R^{}\bigr)^c \end{array} \right)
\,\,+\,\, {\rm H.c.} ~,
\end{eqnarray}
where  $\nu_L^{}$ and $N_R^{}$ are column matrices containing $\nu_{iL}^{}$ and $N_{iR}^{}$,
respectively, and
\begin{eqnarray}
M_{\rm seesaw}^{} \,\,=\,\,
\left( \begin{array}{cc} 0 & m_D^{\rm T} \vspace{0.5ex} \\ m_D^{} & M_N^{}
\end{array} \right) ,
\end{eqnarray}
with the Dirac mass matrix  \,$m_D^{}=v Y_D^{}/\sqrt2$.\,

One can relate the weak eigenstates $\nu_{iL}^{}$ and $(N_{iR}^{})^c$ to the corresponding
mass eigenstates by writing
\begin{eqnarray} \label{basis}
\left( \begin{array}{c} \nu_L^{} \vspace{1ex} \\ \bigl(N_R^{}\bigr)^c \end{array} \right)
=\,\, U \left( \begin{array}{c} \nu_{mL}^{} \vspace{1ex} \\ N_{mL}^{} \end{array} \right) ,
\hspace{5ex}
U \,\,\equiv\,\, \left(\begin{array}{cc} U_{\nu\nu}^{} & U_{\nu N}^{} \vspace{1ex} \\
U_{N \nu}^{} & U_{NN}^{} \end{array}\right) .
\end{eqnarray}
where  $\nu_{mL}^{}$ and $N_{mL}^{}$ are column matrices containing the mass eigenstates.
Thus $U$ is unitary and diagonalizes $M_{\rm seesaw}$,
\begin{eqnarray} \label{mdiag}
\left(\begin{array}{cc} \hat m_\nu^{} & 0 \vspace{0.5ex} \\ 0 & \hat M_N^{} \end{array}\right)
=\,\, U^{\rm T}  M_{\rm seesaw}^{} U ~,
\end{eqnarray}
where  \,$\hat m_\nu^{}={\rm diag}\bigl(m_{\nu_1},m_{\nu_2},m_{\nu_3}\bigr)$\,  and
\,$\hat M_N^{} = {\rm diag}\bigl(M_1, M_2, M_3\bigr)$.\,
On the other hand, the submatrices  $U_{\nu\nu}$, $U_{\nu N}$, $U_{N \nu}$, and $U_{NN}$
are not unitary.
Assuming that the nonzero elements of $M_N^{}$ are all much greater than those of $m_D^{}$,
and expanding in terms of $m_D^{}M_N^{-1}$, one then finds to leading order that $U_{\nu\nu}^{}$
has small deviations from unitarity,  \,$U_{\nu N}^{}=m_D^\dagger\hat M_N^{-1}$,\,
\,$U_{N\nu}^{}=-M_N^{-1}m_D^{}U_{\nu\nu}^{}$,\, \,$U_{NN}^{}=1$,\, and the reduced
light-neutrino mass matrix \,$m_\nu^{}\equiv-m_D^\dagger\hat M_N^{-1}m_D^*$,\,
which can be diagonalized using the unitary Pontecorvo-Maki-Nakagawa-Sakata
matrix $U_{\rm PMNS}$~\cite{pmns}, \,$\hat m_\nu^{}=U_{\rm PMNS}^\dagger m_\nu^{}U_{\rm PMNS}^*$.\,
This leads to the leading-order relation
\begin{eqnarray} \label{UmU}
U_{\rm PMNS}^{}\, \hat m_\nu^{}\, U_{\rm PMNS}^{\rm T} \,\,=\,\,
-U_{\nu N}^{}\, \hat M_N^{}\, U_{\nu N}^{\rm T} ~.
\end{eqnarray}

In terms of the weak eigenstates, the neutrinos couple to the gauge and Higgs bosons in
the~SM according to
\begin{eqnarray}
{\cal L}' \,\,=\,\,
\Biggl( \frac{g}{\sqrt2}\, \bar l_L^{}\, \gamma^\mu \nu_L^{} W^-_\mu
- \bar N_R^{}\, m_D^{} \nu_L^{}\, \frac{h}{v} \,+\, {\rm H.c.} \Biggr)
\,+\, \frac{g}{2 c_{\rm w}^{}}\, \bar\nu_L^{} \gamma^\mu \nu_L^{} Z_\mu^{} ~,
\end{eqnarray}
where \,$g=2m_W^{}/v$\, is the usual weak coupling constant and
\,$c_{\rm w}^{}=\cos\theta_{\rm W}^{}$.\,
Using the relations  \,$U_{N\nu}^{\rm T} m_D^{}=\hat m_\nu^{} U_{\nu\nu}^\dagger$\,
and  \,$U_{NN}^{\rm T} m_D^{}=\hat M_N^{} U_{\nu N}^\dagger$\,  derived from
Eq.~(\ref{mdiag}), one can rewrite ${\cal L}'$ in the mass-eigenstate basis as
\begin{eqnarray} \label{L'}
{\cal L}' &\!\!=&\!\!
\frac{g}{\sqrt2} \Bigl(
\bar l_L^{}\, \gamma^\mu U_{\nu\nu}^{} \nu_{mL}^{} W^-_\mu +
\bar l_L^{}\, \gamma^\mu U_{\nu N}^{} N_{mL} ^{} W^-_\mu \,+\, {\rm H.c.} \Bigr)
\nonumber \\ && +\,\,
\frac{g}{2 c_{\rm w}^{}} \Bigl(
\bar\nu_{mL}^{}\, \gamma^\mu U^\dagger_{\nu\nu} U_{\nu\nu}^{} \nu_{mL}^{}
+ \bar N_{mL}^{}\, \gamma^\mu U_{\nu N}^\dagger U_{\nu\nu}^{} \nu_{mL}^{}
\nonumber \\ && \hspace*{8ex} +\,\,
\bar\nu_{mL}^{}\, \gamma^\mu U^\dagger_{\nu\nu} U_{\nu N}^{} N_{mL}^{}
+ \bar N_{mL}^{}\, \gamma^\mu U_{\nu N}^\dagger U_{\nu N}^{} N_{mL}^{} \Bigr) Z_\mu^{}
\nonumber\\ && -\,\, \Bigl[
\overline{\bigl(\nu_{mL}^{}\bigr)^{\!c}}\,\hat m_\nu^{}U_{\nu\nu}^\dagger U_{\nu\nu}^{}\nu_{mL}^{}
+ \overline{\bigl(N_{mL}^{}\bigr)^{\!c}}\,\hat M_N^{} U_{\nu N}^\dagger U_{\nu\nu}^{} \nu_{mL}^{}
\nonumber \\ && \hspace*{4ex} +\,\,
\overline{\bigl(\nu_{mL}^{}\bigr)^{\!c}}\,\hat m_\nu^{} U_{\nu\nu}^\dagger U_{\nu N}^{} N_{mL}^{}
+ \overline{\bigl(N_{mL}^{}\bigr)^{\!c}}\,\hat M_N^{} U_{\nu N}^\dagger U_{\nu N}^{} N_{mL}^{}
\,+\, {\rm H.c.} \Bigr] \frac{h}{v} ~.
\end{eqnarray}
Thus via mixing the heavy neutrinos $N$ can interact with the SM gauge bosons at tree level.

This Lagrangian indicates that the Higgs-boson coupling to a pair of light and
heavy neutrinos, $h\nu N$, is leading compared to the other Higgs-neutrino couplings,
$h\nu\nu$ and $hNN$, which are proportional to the tiny light-neutrino masses and
of second order in~$U_{\nu N}$, respectively.
This dominant coupling generates the decay mode \,$h\to\nu N$\, if the mass of the heavy
neutrino is less than the Higgs mass~$m_h^{}$.
Clearly, how important this decay might be would depend on the elements of
the matrix~$U_{\nu N}$, which parametrizes the light-heavy mixing.
As we will show later, the elements of $U_{\nu N}$ subject to current experimental constraints
can be sufficiently sizable to give rise to significant modifications of the Higgs decay
branching ratios in the~SM for a~relatively light Higgs-boson.

\subsection{\boldmath$h\to\nu N$\, decay}

From Eq.~(\ref{L'}), we obtain the amplitude for \,$h\to \nu_i^{}N_j^{}$
\begin{eqnarray} \label{M_h2nN}
{\cal M}\bigl(h\to\nu_i^{} N_j\bigr) \;=\; \frac{g\,M_j^{}}{2m_W^{}}\,
\bar u_\nu^{} \Bigl[ \bigl(U_{\nu\nu}^{\rm T}U_{\nu N}^*\bigr)_{ij} P_{\rm L}^{}
+ \bigl(U_{\nu\nu}^\dagger U_{\nu N}^{}\bigr)_{ij} P_{\rm R}^{} \Bigr] v_N^{} ~,
\end{eqnarray}
where $\nu_i^{}$ and $N_i^{}$ denote the $i$th mass-eigenstates of the light and
heavy neutrinos, respectively, $M_i^{}$ is the mass of $N_i^{}$, and
\,$P_{\rm L,R}^{}=\frac{1}{2}(1\mp\gamma_5^{})$.\,
In deriving this expression, we have made use of the Majorana nature of both neutrinos and
neglected contributions from terms in ${\cal L}'$ proportional to the light-neutrino masses.
The resulting decay rate for all possible combinations of $\nu_i^{}N_j^{}$ is
\begin{eqnarray} \label{w_h2nN}
\Gamma(h\to\nu N) \;=\; \sum_{i,j=1}^3 \Gamma\bigl(h\to\nu_i^{}N_j\bigr) \;=\;
\sum_{i=1}^3\frac{g^2 m_h^{}M_i^2\, \bigl(U_{\nu N}^\dagger U_{\nu N}^{}\bigr)_{ii}}{32\pi\,m_W^2}
\Biggl(1-\frac{M_i^2}{m_h^2}\Biggr)^{\!\!2} ~,
\end{eqnarray}
with \,$U_{\nu\nu}^{}U_{\nu\nu}^\dagger\simeq1$.\,
For our numerical analysis, we will employ some of the results of Ref.~\cite{He:2009ua} which
provided specific solutions for $U_{\nu N}$ having sizable elements and simultaneously
satisfying the light-neutrino mass requirement given in Eq.~(\ref{UmU}).

There are additional sets of constraints that the elements of $U_{\nu N}$ must satisfy.
The first arises from electroweak precision data on processes involving neutral currents
conserving lepton flavor~\cite{pdg}.
Expressed in terms of \,$\epsilon\equiv U_{\nu N}^{}U_{\nu N}^\dagger$,\, in type-I seesaw
the bounds  extracted from the data are~\cite{delAguila:2008cj,ewpd}
\begin{eqnarray} \label{eii}
\epsilon_{11}^{} \,\,\le\,\, 3.0\times10^{-3} ~, \hspace{5ex}
\epsilon_{22}^{} \,\,\le\,\, 3.2\times10^{-3} ~, \hspace{5ex}
\epsilon_{33}^{} \,\,\le\,\, 6.2\times10^{-3} ~.
\end{eqnarray}
The second set of constraints come from lepton-flavor violating transitions in
the charged-lepton sector.
Although in type-I seesaw there are no flavor-changing processes involving ordinary charged
leptons at tree level, loop-induced ones can occur, such as the radiative decays
\,$\mu\to e\gamma$,\, \,$\tau\to e\gamma$,\, and \,$\tau\to\mu\gamma$.\,
The bounds determined from the measurements of such transitions are~\cite{ewpd,typeI_fcnc}
\begin{eqnarray} \label{eij}
|\epsilon_{12}^{}| \,\,\le\,\, 1\times10^{-4} ~, \hspace{5ex}
|\epsilon_{13}^{}| \,\,\le\,\, 0.01 ~, \hspace{5ex} |\epsilon_{23}^{}| \,\,\le\,\, 0.01 ~.
\end{eqnarray}
For heavy neutrinos coupling to the electron, neutrinoless double-beta decay
imposes~\cite{Belanger:1995nh}
\begin{eqnarray} \label{ee0nu}
\biggl|\sum_{i=1}^3(U_{\nu N})_{1i}^2/M_i^{}\biggr| \,\,\le\,\,
5\times10^{-8}{\rm\,GeV}^{-1} ~.
\end{eqnarray}
Finally, for $N$-mass values between a few GeV and the $Z$ mass, $m_Z^{}$, there are also
restrictions on the individual elements $(U_{\nu N})_{2i}$ and $(U_{\nu N})_{3i}$ from
searches for SM-singlet neutrinos via \,$Z\to\nu N$\, performed by the L3 and DELPHI
experiments at LEP~\cite{lep}.
These constraints on $(U_{\nu N})_{2i,3i}$ may be stronger than those inferred from
Eqs.~(\ref{eii}) and~(\ref{eij}), depending on~$M_i$.

To explore the effect of large light-heavy mixing on the decay \,$h\to\nu N$,\, we take some
of the examples of $U_{\nu N}$ from Ref.~\cite{He:2009ua}.
As discussed therein, the general form of $U_{\nu N}$ which accommodates the large mixing
can be written as \,$U_{\nu N}=U_0+U_\delta$,\, where $U_0$ is a rank-one matrix which makes
the right-hand side of Eq.~(\ref{UmU}) vanish exactly and $U_\delta$ denotes a perturbation
matrix with tiny elements fixed to reproduce the light-neutrino masses according to
Eq.~(\ref{UmU}).
It follows that the elements of $U_0$ are not constrained by the light-neutrino masses and
can be as large as allowed by the experimental bounds described in the preceding paragraph.
We note that, as mentioned earlier, this situation can happen naturally in the presence of
some underlying symmetry that is slightly violated~\cite{Kersten:2007vk}.
Accordingly, in our discussion below we include in $U_{\nu N}$ only the dominant
part,~\,$U_{\nu N}=U_0$.

\begin{figure}[t]
\includegraphics[width=168mm]{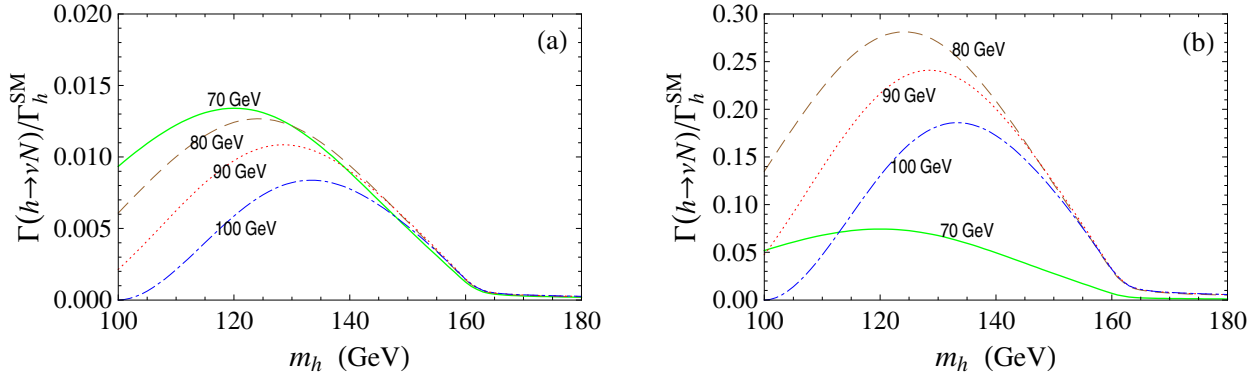} \vspace*{-2ex}
\caption{Ratios of the width of \,$h\to\nu N$\, in type-I seesaw to the total Higgs width in
the SM as functions of the Higgs mass $m_h^{}$ for heavy-neutrino mass values
\,$m_N^{}=70,80,90,100$~GeV\, and different choices of $U_{\nu N}^{}$ as described in
the text.\label{ratio1}} \vspace*{1ex}
\end{figure}

Thus, for the first example, we take~\cite{He:2009ua}
\begin{eqnarray} \label{U0a}
U_{\nu N}^{} \,\,=\,\, U_0^a \,\,= \left( \begin{array}{lll}
a \,&\, a \,&\, i \sqrt2\, a \vspace{0.5ex} \\ b \,&\, b \,&\, i\sqrt2\, b \vspace{0.5ex} \\
c \,&\, c \,&\, i\sqrt2\, c \end{array}\right) \!{\cal R} ~, \hspace{5ex}
{\cal R} \,\,=\,\, {\rm diag}\Bigl(\sqrt{r_1^{}},\,\sqrt{r_2^{}},\,\sqrt{r_3^{}}\Bigr) ~,
\end{eqnarray}
where  \,$a=(0.58-0.81\,i)\bar b$,\, \,$b=(0.58+0.41\,i)\bar b$,\, \,$c=(0.58+0.41\,i)\bar b$,\,
and \,$r_i^{}=m_N^{}/M_i^{}$,\, with $\bar b$ being a free parameter that has to satisfy
the bounds listed above and $m_N^{}$ taken to be the lightest of $M_{1,2,3}$.
From now on, for simplification we assume that the heavy neutrinos are degenerate,
\,$M_1^{}=M_2^{}=M_3^{}=m_N^{}$,\, the situation in the nondegenerate case being qualitatively
similar, and so $\cal R$ is a unit matrix in the following examples, but not explicitly displayed.
We then obtain \,$\bar b=0.006$\, to be the largest value allowed by the experimental constraints.
Adopting this number, we plot in Fig.~\ref{ratio1}(a) the ratio of $\Gamma(h\to\nu N)$ to
the total Higgs width $\Gamma_h^{\rm SM}$ in the SM corresponding to Higgs mass values within
the range \,$100{\rm\,GeV}\le m_h^{}\le180{\rm\,GeV}$\, for \,$m_N^{}=70,80,90,100$~GeV.\,
From the curves displayed, the peak of the ratio is seen to be only about \,$1.3\%$,\,
corresponding to the \,$m_N^{}=70$\,GeV\, curve at \,$m_h^{}=120$\,GeV.\,
We remark that for lower values of $m_N^{}$, from several to 60~GeV,\, the LEP searches
mentioned earlier impose the strong limits
\,$|(U_{\nu N})_{2i,3i}|\mbox{\footnotesize\,$\lesssim$\,}0.007$\,~\cite{lep}.

As a second example, we can choose~\cite{He:2009ua}
\begin{eqnarray} \label{U0d}
U_{\nu N}^{} \,\,=\,\, U_0^d \,\,= \left( \begin{array}{ccc} 0 \,&\, a \,&\, i a \vspace{0.5ex} \\
0 & b & i b \vspace{0.5ex} \\ 0 & c & i c \end{array}\right)
\end{eqnarray}
with  \,$a=-0.82\,\bar a$,\, \,$b=(0.41+0.66\,i)\bar a$,\, and \,$c=(0.41-0.66\,i)\bar a$,\,
where $\bar a$ is a free parameter subject to the experimental constraints.
We find that the maximum allowed value \,$\bar a=0.0089$\, leads to a graph very similar to
that in Fig.~\ref{ratio1}(a).
It is worth noting that $U_{\nu N}^{}$ in either Eq.~(\ref{U0a}) or~(\ref{U0d}) automatically
satisfies the constraint in Eq.~(\ref{ee0nu}) for degenerate heavy neutrinos.

It is evident that the effect of light-heavy mixing on \,$h\to\nu N$\, in the preceding examples
is not remarkable.
This is mainly because of the strict bound on $\epsilon_{12}^{}$ in Eq.~(\ref{eij}) which
limits the elements of $U_{\nu N}^{}$ to be at most~$\sim$0.01\, in size.
It turns out that there are other choices of~$U_{\nu N}^{}$ which can evade this restriction,
one of them being~\cite{He:2009ua}
\begin{eqnarray} \label{u0e}
U_{\nu N}^{} \,\,=\,\, U_0^e \,\,= \left(\begin{array}{ccc} 0 & 0 & 0 \vspace{0.5ex} \\
0 \,&\, a \,&\, ia \vspace{0.5ex} \\ 0 & b & ib \end{array}\right)
\end{eqnarray}
with  \,$b=a$.\,
We obtain the largest allowed value of $a$ to be \,$a=0.04$\, for
\,$m_N^{}\,\raisebox{-0.4ex}{\footnotesize$\stackrel{\textstyle>}{\sim}$}\,80$\,GeV,\,
but \,$a\sim0.02$\, for \,$m_N^{}=70$\,GeV\, from the LEP searches~\cite{lep}.
These numbers lead to \,$h\to\nu N$\, rates which are much larger than those in the earlier
examples, and the ratios of these enlarged rates to the SM Higgs total width are shown in
Fig.~\ref{ratio1}(b).
More precisely, one observes from the four curves displayed that the ratio can reach as high as
\,28\%,\, which corresponds to the peak of the \,$m_N^{}=80$\,GeV\, curve at \,$m_h^{}=124$\,GeV.\,
Another choice that can evade the $\epsilon_{12}^{}$ constraint is~\cite{He:2009ua}
\begin{eqnarray} \label{u0f}
U_{\nu N}^{} \,\,=\,\, U_0^f \,\,= \left( \begin{array}{ccc} 0 \,&\, a \,&\, i a \vspace{0.5ex} \\
0 & 0 & 0 \vspace{0.5ex} \\ 0 & b & i b \end{array}\right)
\end{eqnarray}
with \,$b=(0.0013+1.03i)a$\, and \,$a=0.02$~(0.039)\, for \,$m_N^{}=70$\,GeV~$(\ge80$\,GeV).\,
This results in a plot very similar to that in Fig.~\ref{ratio1}(b).

Since the light-heavy mixing also causes the $Z$ coupling to $\nu N$, according to
Eq.~(\ref{L'}), the decay \,$Z\to\nu N$\, can happen for \,$m_N^{}<m_Z^{}$.\,
Therefore, it is important to check if the impact on \,$Z\to\nu N$\, from the large mixing
we are considering is consistent with the precisely measured value of the $Z$ total-width,
\,$\Gamma_Z^{\rm exp}=2.4952\pm0.0023$\,GeV\,~\cite{pdg}, which agrees with the standard-model
prediction \,$\Gamma_Z^{\rm SM}=2.4954\pm0.0010$\,GeV\,~\cite{Erler:2009jh}.
From Eq.~(\ref{L'}), one derives the amplitude
\begin{eqnarray}
{\cal M}\bigl(Z\to\nu_i^{}N_j^{}\bigr) \,\,=\,\,
\frac{g\,\varepsilon_Z^\mu}{2c_{\rm w}^{}}\, \bar u_\nu^{} \gamma_\mu^{}\Bigl[
\bigl(U_{\nu\nu}^{\rm T}U_{\nu N}^*\bigr)_{ij}P_{\rm R}^{}
- \bigl(U_{\nu\nu}^\dagger U_{\nu N}^{}\bigr)_{ij}P_{\rm L}^{} \Bigr] v_N^{} ~,
\end{eqnarray}
upon using the Majorana nature of the neutrinos.
One then gets the decay rate
\begin{eqnarray}
\Gamma(Z\to\nu N) \,\,=\,\, \sum_{i,j=1}^3\Gamma\bigl(Z\to\nu_i^{}N_j^{}\bigr) \,\,=\,\,
\sum_{i=1}^3 \frac{g^2\,\bigl(U_{\nu N}^\dagger U_{\nu N}^{}\bigr)_{ii}\, m_Z^3}{48\pi\,m_W^2}
\Biggl(1-\frac{3 m_N^2}{2 m_Z^2}+\frac{m_N^6}{2m_Z^6}\Biggr) ~,
\end{eqnarray}
with \,$U_{\nu\nu}^{}U_{\nu\nu}^\dagger\simeq1$\, as before.
Numerically, the choices \,$U_{\nu N}^{}=U_0^e$ and $U_0^f$\, producing the largest effects
obtained above yield nearly identical rates:
\,$\Gamma(Z\to\nu N)\simeq0.12$, 0.16, and 0.002~MeV\, for
\,$m_N^{}=70$, 80, and 90~GeV,\, respectively.
Obviously, each of these $\Gamma(Z\to\nu N)$ numbers is well within the errors in
$\Gamma_Z^{\rm exp}$ and~$\Gamma_Z^{\rm SM}$.
This helps to confirm that our parameter choices for $U_{\nu N}$ already satisfied
the LEP and other constraints described earlier in this subsection.

Another process which should be examined if the large light-heavy mixing occurs is
the scattering \,$e^+e^-\to\nu N$\, followed by the decay \,$N\to\nu l_1^+l_2^-$,\,
where the charged leptons $l_{1,2}^{}$ can be equal or different in flavor.
For this has the same leptonic final-state as the $W$-pair production process
\,$e^+e^-\to WW$,\, each of the $W$'s subsequently decaying into $\nu l$,
which has been well measured at LEP2~\cite{Abdallah:2003zm,Abbiendi:2007rs},
its cross-section found to be in accord with the SM expectation.
The amplitude for \,$e^+e^-\to\nu N$\, is related by crossing symmetry to that for
\,$N\to\nu l^+l^-$,\, to be evaluated later, in Eq.~(\ref{N2nll}) and proceeds from
an $s$-channel $Z$-mediated diagram plus $W$-mediated diagrams in the $t$ and $u$ channels.
For completeness, here we write its squared amplitude as
\begin{eqnarray}
\overline{\bigl|{\cal M}\bigl(e^+e^-\to\nu_i^{}N_j^{}\bigr)\bigr|^2} &=&
\frac{g^4\,\bigl(l_e^2+r_e^2\bigr)}{4 c_{\rm w}^4}
\Bigl|\bigl(U_{\nu\nu}^\dagger U_{\nu N}^{}\bigr)_{ij}\Bigr|^2\,
\frac{\bigl(t-m_N^2\bigr)t+\bigl(u-m_N^2\bigr)u}{\bigl(s-m_Z^2\bigr)^2}
\nonumber \\ && \!\! +\;
\frac{g^4}{4}\Bigl|\bigl(U_{\nu\nu}^\dagger\bigr)_{i1}\bigl(U_{\nu N}^{}\bigr)_{1j}\Bigr|^2
\left[\frac{\bigl(u-m_N^2\bigr)u}{\bigl(t-m_W^2\bigr)^2} +
\frac{\bigl(t-m_N^2\bigr)t}{\bigl(u-m_W^2\bigr)^2}\right]
\nonumber \\ && \!\! +\;
\frac{g^4 l_e^{}\,{\rm Re}\Bigl[\bigl(U_{\nu\nu}^\dagger U_{\nu N}^{}\bigr)_{ij}
\bigl(U_{\nu\nu}^{}\bigr)_{1i}\bigl(U_{\nu N}^\dagger\bigr)_{j1}\Bigr]}
{2c_{\rm w}^2\,\bigl(s-m_Z^2\bigr)}\Biggl[
\frac{\bigl(u-m_N^2\bigr)u}{t-m_W^2} +
\frac{\bigl(t-m_N^2\bigr)t}{u-m_W^2} \Biggr] ~, ~~~~~~~
\end{eqnarray}
where we have neglected light-lepton masses and $\Gamma_{W,Z}$ terms,
\,$s=\bigl(p_{e^+}^{}+p_{e^-}^{}\bigr){}^2$,\, \,$t=\bigl(p_{e^+}^{}-p_N^{}\bigr){}^2$,\,
\,$u=m_N^2-s-t$,\, \,$l_e^{}=s_{\rm w}^2-\frac{1}{2}$,\,
and \,$r_e^{}=s_{\rm w}^2$,\, with \,$s_{\rm w}^{}=\sin\theta_{\rm W}^{}$.\,
We sum this over all possible $\nu_i^{}N_j^{}$ combinations and then,
in order to make comparison with the LEP2 data, apply the resulting cross-section in
\,$\sigma\bigl(e^+e^-\to\nu N\to\nu l_1^+\nu l_2^-\bigr)=
\sigma(e^+e^-\to\nu N)\,{\cal B}\bigl(N\to\nu l_1^+l_2^-\bigr)$,\,
employing \,${\cal B}\bigl(N\to\nu l_1^+l_2^-\bigr)\sim0.3$,\,
as we will calculate in the next subsection.
We have collected the numbers in Table~\ref{cs} for \,$m_N^{}=70$,\,80,\,90,\,and\,100~GeV,\,
with \,$U_{\nu N}^{}=U_0^e$ and $U_0^f$\, as before, at the center-of-mass energies
\,$\sqrt s=161$, 183, and 207~GeV,\, which are representative of the measured range.
As one can notice from the table, in this case the impact of $U_0^e$ is much smaller
than that of $U_0^f$, which is due to the fact that \,$\bigl(U_0^e\bigr){}_{1j}^{}=0$.\,
The experimental cross-sections of \,$e^+e^-\to WW\to\nu l_1^+\nu l_2^-$\, reported by the LEP2
collaborations are consistent with each other~\cite{Abdallah:2003zm,Abbiendi:2007rs},
and so it suffices to compare with the most recent ones, from OPAL~\cite{Abbiendi:2007rs},
which we have reproduced in Table~\ref{cs}, after combining the statistical and systematic
errors in quadrature.
It is evident from this table that all the $\nu N$ contributions, especially the ones
arising from $U_0^e$, are well within the uncertainties in the data.
Hence our large-mixing results are compatible with the LEP2 measurements.

We have thus demonstrated that, with the large light-heavy mixing subject to current
experimental constraints, the new decay mode \,$h\to\nu N$\, can change the total Higgs
width in the SM by up to nearly 30\% for a relatively light Higgs-boson,
especially with \,$m_h^{}\mbox{\footnotesize\,$\lesssim$\,}140$\,GeV.\,
This would significantly affect the SM expectations in Higgs searches at the LHC.
For bigger Higgs masses, \,$m_h^{}>2m_W^{}$,\, as the decay channels into a pair of weak gauge
bosons,\, $h\to WW,ZZ$,\, become open and start to be dominant, the effect of
\,$h\to\nu N$\, due to the large mixing on the Higgs total width would get much reduced,
as can be seen also from Fig.~\ref{ratio1}.
We can mention here that the possibility of \,$h\to\nu N$\, causing important changes to
Higgs searches in the presence of the large mixing has also been raised previously in
Ref.~\cite{Pilaftsis:1991ug} which proposed radiatively induced neutrino masses and
more recently in Ref.~\cite{de Gouvea:2007uz} in the contexts of other models of
light-neutrino mass generation.

Since \,$h\to\nu N$\, is a potentially influential decay mode, it is of interest as well
to study the subsequent decays of $N$ in the case of large light-heavy mixing.
They may have signatures which are observable and distinguishable from those of the~SM.
We explore this possibility in the rest of this section.

\begin{table}[b]
\caption{Cross-section of \,$e^+e^-\to\nu N\to\nu l_1^+\nu l_2^-$\, for
\,$m_N^{}=70,80,90,100$~GeV\, with \,$U_{\nu N}^{}=U_0^e$\, in Eq.\,(\ref{u0e}) (columns 2-5)\,
and  \,$U_{\nu N}^{}=U_0^f$\, in Eq.\,(\ref{u0f}) (columns 6-9),\,
compared to measured cross-section of \,$e^+e^-\to WW\to\nu l_1^+\nu l_2^-$\, (last column)
at center-of-mass energies \,$\sqrt s=161,183,207$\,GeV.
All~cross-section numbers are in~pb.}
\medskip
\begin{tabular}{|c||cccc|cccc|c|}
\hline
\multirow{2}{*}{\footnotesize\, $\sqrt s$ (GeV) \,} &
\multicolumn{4}{c|}{\scriptsize$m_N^{}\vphantom{\int_|^|}$} &
\multicolumn{4}{c|}{\scriptsize$m_N^{}\vphantom{\int_|^|}$} & \multirow{2}{*}{Data~\cite{Abbiendi:2007rs}} \\
\cline{2-9} & \, {\scriptsize70\,GeV}$\vphantom{\int_|^|}$ \, & \, {\scriptsize80\,GeV} \, &
\, {\scriptsize90\,GeV} \, & \, {\scriptsize100\,GeV} \, & \,
{\scriptsize70\,GeV}$\vphantom{\int_|^|}$ \, & \, {\scriptsize80\,GeV} \, &
\, {\scriptsize90\,GeV} \, & \, {\scriptsize100\,GeV}  \, & $$ \\
\hline\hline
161 & 0.001 & 0.005 & 0.005 & 0.004 & 0.013 & 0.047 & 0.050 & 0.040 & \, $0.28\pm0.22$ \, \\
183 & 0.001 & 0.003 & 0.003 & 0.003 & 0.014 & 0.052 & 0.057 & 0.048 & \, $1.63\pm0.21$ \, \\
207 & 0.001 & 0.002 & 0.003 & 0.002 & 0.014 & 0.056 & 0.063 & 0.055 & \, $1.83\pm0.13$ \,\\
\hline
\end{tabular} \label{cs}
\end{table}

\subsection{Leading decays of \boldmath$N$}

For \,$m_N^{}<m_W^{}$,\, the dominant decay-modes of $N$ are into three light fermions.
If  \,$m_h^{}>m_N^{}>m_W^{}$ or~$m_Z^{}$,\, the main $N$ decays are effectively still
three-body, as the daughter $W$ or $Z$ promptly decays into a pair of light fermions.
We have derived the amplitudes for these decays and collected their expressions in the appendix.

We have computed the corresponding decay rates for \,$m_N^{}=70,80,90,100$~GeV\, and
\,$U_{\nu N}^{}=U_0^e$\, in Eq.~(\ref{u0e}), with the same \,$b=a$ values as those chosen for
Fig.~\ref{ratio1}(b).
Moreover, we have summed the rates over all possible final-states, taking into account
the number of colors in final states involving quarks and excluding top-quark contributions.
The results are listed in Table~\ref{rates}, where \,$l'\neq l$\, in the second decay mode.
With \,$U_{\nu N}^{}=U_0^f$\, in Eq.~(\ref{u0f}) instead, we get similar numbers.
One can observe in this table that there are large increases in the rates of some of the modes
between \,$m_N^{}=80$ and $90$~GeV\,  or between \,$m_N^{}=90$ and $100$~GeV,\, which are to be
expected due to the opening of decay channels with an on-shell $W$ or $Z$.

\begin{table}[b]
\caption{Rates, in keV, of $N$ decays into three light fermions for \,$m_N^{}=70,80,90,100$~GeV\,
and \,$U_{\nu N}^{}=U_0^e$\, in Eq.~(\ref{u0e}), with the same \,$b=a$ values as for
Fig.~\ref{ratio1}(b).}
\medskip
\begin{tabular}{|l|cccc|}
\hline
\multirow{2}{*}{\, \, Decay mode} &
\multicolumn{4}{c|}{$m_N^{}\vphantom{\int_|^|}$} \\
\cline{2-5} & \, 70\,GeV$\vphantom{\int_|^|}$ \, & \, \, 80\,GeV \, & \, \, 90\,GeV \, &
\, 100\,GeV \, \\
\hline\hline
\, \, \, $N\to\nu\nu\bar\nu\vphantom{\int_|^|}$      & 0.06 & \, 0.6 &  \, \, 2 & \, 16 \\
\, \, \, $N\to\nu l^+l^{\prime-}\vphantom{\int_|^|}$ & 0.24 & \, 3.4 &  \,   38 &   128 \\
\, \, \, $N\to\nu l^+l^-\vphantom{\int_|^|}$         & 0.11 & \, 1.5 &  \,   19 & \, 71 \\
\, \, \, $N\to\nu q\bar q\vphantom{\int_|^|}$        & 0.25 & \, 2.4 &  \, \, 7 & \, 63 \\
\, \, \, $N\to l^- u\bar d\vphantom{\int_|^|}$       & 0.37 & \, 5.1 &  \,   57 &   192 \\
\, \, \, $N\to l^+\bar u d\vphantom{\int_|^|}$       & 0.37 & \, 5.1 &  \,   57 &   192 \\
\hline
\, $N\to3~\rm fermions\vphantom{\int_{\big|}^{\big|}}$ \, & 1.40 &   18.1 &  180 &   662 \\
\hline
\end{tabular} \label{rates} \vspace*{-1ex}
\end{table}

From the entries in the last row of Table~\ref{rates}, we can estimate the total widths of $N$
for the different $m_N^{}$ values, namely  \,$\Gamma_N^{}\simeq\Gamma(N\to3\rm\,fermions)$.\,
For any one of these $m_N^{}$ values, we can then determine how far $N$ is likely to travel
after being produced in the Higgs decay \,$h\to\nu N$\, and before decaying in the rest frame
of~$h$, once $m_h^{}$ is specified.
We have found that in this case the largest distance traveled by $N$ is less than
\,$10^{-10}$\,m\, for the $m_h^{}$ values considered here, namely
\,$100{\rm\,GeV}\le m_h^{}\le180{\rm\,GeV}$.\,
More specifically, in the rest frame of the decaying Higgs boson, the most energetic and
longest-lived $N$ corresponds to  \,$m_h^{}=180$\,GeV,\, \,$m_N^{}=70$\,GeV,\, and
\,$\Gamma_N^{}\simeq3$\,keV,\, as Table~\ref{rates} indicates, and its maximum distance is
calculated to be \,$d_N^{}\simeq7\times10^{-11}$\,m.\,

These considerations imply that the decay \,$h\to\nu N$\, is very quickly followed by $N$
decaying into three light fermions, and hence this decay sequence is effectively a four-body
Higgs decay, \,$h\to\nu ff'f''$,\, each $f$ being a light fermion.
It is interesting to compare these $N$-mediated Higgs decays with their counterparts
in the~SM, which arise mostly from diagrams mediated by a pair of $W$ or $Z$ bosons, as well as
with the other Higgs decay modes in the~SM.

\begin{figure}[b] \vspace*{1ex}
\includegraphics[width=168mm]{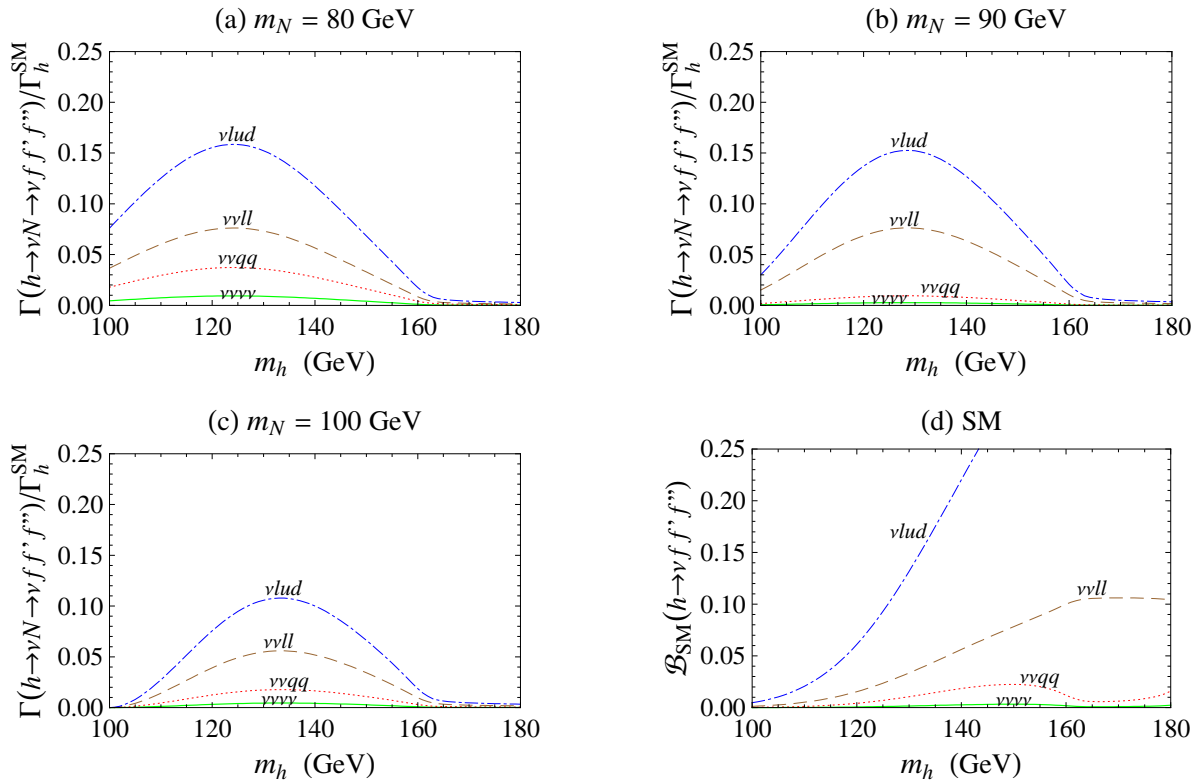}
\caption{(a,b,c) Ratios of the widths of \,$h\to\nu N\to\nu ff'f''$ in type-I seesaw to the total
Higgs width in the SM as functions of the Higgs mass $m_h^{}$ for heavy-neutrino mass values
\,$m_N^{}=80,90,100$~GeV\, and \,$U_{\nu N}^{}=U_0^e$\, in Eq.~(\ref{u0e}), with the same
\,$b=a$\, value as for Fig.~\ref{ratio1}(b).\,
(d)~Branching ratios of \,$h\to\nu ff'f''$\, in the SM.\label{ratio2}}
\end{figure}

In Fig.~\ref{ratio2}(a,b,c), we display the ratios of the widths of \,$h\to\nu N\to\nu ff'f''$
to the Higgs total width $\Gamma_h^{\rm SM}$ in the SM as functions of the Higgs mass $m_h^{}$
for \,$m_N^{}=80,90,100$~GeV\,  and \,$U_{\nu N}^{}=U_0^e$\, in
Eq.~(\ref{u0e}), with the same \,$b=a$ values as those chosen for Fig.~\ref{ratio1}(b).
For comparison, Fig.~\ref{ratio2}(d) shows the branching ratios of \,$h\to\nu ff'f''$\, in
the~SM, which are induced by \,$h\to WW^{(*)},ZZ^*$\, diagrams.
In each of these four graphs, the curve labeled $\nu\nu\nu\nu$ corresponds to the rates of
the $\nu\nu\bar\nu\bar\nu$ modes,  \,$\nu\nu ll$\, to the combined rates of
the \,$\nu\bar\nu l^+l^{\prime-}$\, and  \,$\nu\bar\nu l^+l^-$\, modes, \,$\nu\nu qq$\, to
the rates of the \,$\nu\bar\nu q\bar q$\, modes, and \,$\nu lud$\, to the combined rates of
the \,$\nu l^+\bar u d$\, and  \,$\bar\nu l^-u\bar d$\, modes.

Evidently, for $m_h^{}$ less than 140\,GeV or so, the $N$-mediated contributions to each of
the four-body modes graphed in Fig.~\ref{ratio2} are comparable to,
and can be a few times bigger than, the corresponding SM contributions.
This is clearly the case when it comes to the $\nu lud$ and $\nu\nu ll$ curves for
the three values of $m_N^{}$ considered.
We remark that in the plots (a,b,c) we have not included possible interference between
the $N$-mediated and SM contributions, but it should be taken into account in a more refined
analysis.
Nevertheless, this exercise serves to demonstrate the potential importance of the effect of
large light-heavy mixing on Higgs decays.
Accordingly, if the Higgs boson is detected,
with \,$m_h\mbox{\footnotesize\,$\lesssim$\,}140$\,GeV,\, and its decay modes can be studied
with sufficient precision, these four-body Higgs decays may offer useful information on
the seesaw mechanism.
This information would be complementary to that possibly available from direct searches
for $N$ at colliders, such as via \,$pp\to W^*X\to l N X$\, at
the LHC~\cite{He:2009ua,delAguila:2008cj}.
Lastly, it is worth pointing out that, as the $\nu lud$ and $\nu\nu ll$ curves in
Fig.~\ref{ratio2}(a,b,c) indicate, the rates of these four-body decays are not much smaller than
the SM rate of \,$h\to b\bar b$\, for the $m_h^{}$ range shown and can be much larger
than the rates of other SM modes, such as \,$h\to c\bar c,gg,l^+l^-$.\,

\section{Large light-heavy mixing in type-III seesaw}

In type-III seesaw the SM-singlet neutrinos in type-I seesaw are replaced by weak-SU$(2)_{\rm L}$
triplets of right-handed heavy leptons having zero hypercharge~\cite{Foot:1988aq}.
The component fields of each triplet $\Sigma$ and its charge conjugate
\,$\Sigma^c=C\bar\Sigma^{\rm T}$\, are
\begin{eqnarray}
\Sigma \,\,= \left(\begin{array}{cc} N^0/\sqrt2 & E^+ \vspace{0.5ex} \\
E^- & -N^0/\sqrt2 \end{array} \right), \hspace{5ex}
\Sigma^c \,\,= \left( \begin{array}{cc} N^{0c}/\sqrt2 & E^{-c} \vspace{0.5ex} \\
E^{+c} & -N^{0c}/\sqrt2 \end{array} \right) ,
\end{eqnarray}
and the renormalizable Lagrangian for each $\Sigma$ is given by
\begin{eqnarray} \label{Liii0}
{\cal L}_{\rm III}^{} \,\,=\,\,
{\rm Tr}\bigl(\bar\Sigma i\!\!\not{\!\!D}\Sigma\bigr)
- \mbox{$\frac{1}{2}$}\, {\rm Tr}\Bigl(\bar\Sigma M_\Sigma^{}\Sigma^c
+ \overline{\Sigma^{{}^{\scriptstyle c}}} M_\Sigma^* \Sigma\Bigr)
- \sqrt2\, \tilde H^\dagger\bar\Sigma Y_\Sigma^{} L_L^{}
- \sqrt2\, \bar L_L^{} Y_\Sigma^\dagger \Sigma \tilde H ~,
\end{eqnarray}
where  $D_\mu$ is a covariant derivative involving the weak gauge bosons, $M_\Sigma^{}$
the mass of the triplet, and $Y_\Sigma$ its Yukawa coupling.
Defining  \,$E=\bigl(E_R^+\bigr){}^c+E_R^-$\, and removing the would-be Goldstone bosons $\eta$
and $\phi^\pm$, one can rewrite ${\cal L}_{\rm III}$ as
\begin{eqnarray} \label{Liii}
{\cal L}_{\rm III}^{} &\!\!=&\!\! \bar E i\!\!\not{\!\partial} E
+ \overline{N_R^0} i\!\!\not{\!\partial} N^0_R - \bar E M_\Sigma^{} E
- \mbox{$\frac{1}{2}$}\Bigl[\overline{N^0_R}M_\Sigma^{}\bigl(N_R^0\bigr)^c\,+\,{\rm H.c.}\Bigr]
\nonumber \\ && \!\! +\,\,
g \left[\overline{N_R^0} \not{\!\!W}^+ E_R^{}
+ \overline{\bigl(N_R^0\bigr)^{\!c}} \not{\!\!W}^+ E_L^{} \,+\, {\rm H.c.} \right]
- g\, \bar E \not{\!\!W}_{\!\!3}^{} E
\nonumber \\ && \!\! -\,\,
\Bigl[ \mbox{$\frac{1}{\sqrt2}$}(v+h)\overline{N_R^0}Y_\Sigma^{}\nu_L^{}
+ (v+h)\bar E Y_\Sigma^{} l_L^{} \,+\, {\rm H.c.} \Bigr] ~,
\end{eqnarray}
where  \,$W_3^\mu=-s_{\rm w}^{}A^\mu+c_{\rm w}^{}Z^\mu$\, is the usual linear combination of
the photon and $Z$-boson fields,  \,$N_R=N$,\, and  \,$E_{L,R}=P_{L,R}E$,\, with
\,$P_{L,R}=\frac{1}{2}(1\mp\gamma_5^{})$.\,

From ${\cal L}_{\rm III}$ in the mass-eigenstate basis, one can then write down the relevant
terms describing the interactions of the heavy leptons $N$ and $E$ with the Higgs boson.
Here we follow the notation of Ref.~\cite{He:2009ua}, where more details on the other terms
in the Lagrangian can be found, and also assume that there are three triplets.
The Higgs couplings of $N$ are the same as those in the type-I seesaw discussed earlier.
The interactions of $E$ are described by
\begin{eqnarray} \label{LE}
{\cal L}_{E}^{} \;=\; \frac{-g}{\sqrt2\,m_W^{}}
\Bigl(\bar l_{mL}^{} U_{\nu N}^{} M_{\Sigma}^{} E_{mR}^{} +
\bar E_{mR}^{} M_{\Sigma}^{} U_{\nu N}^\dagger l_{mL}^{}\Bigr) h  ~+~ \cdots ~,
\end{eqnarray}
where only the relevant part is displayed, $l_{mL}$ and $E_{mR}$ are (3$\times$1) column
matrices containing the mass eigenstates of the light and heavy charged-leptons, respectively,
and $M_\Sigma$ is now a~diagonal matrix, \,$M_\Sigma={\rm diag}\bigl(M_1, M_2, M_3\bigr)$.

The amplitude for \,$h\to\nu N$\, and its decay rate are then those given in
Eqs.~(\ref{M_h2nN}) and~(\ref{w_h2nN}).
For \,$h\to l^-E^+$,\, we have from Eq.~(\ref{LE})
\begin{eqnarray}
{\cal M}\bigl(h\to l_i^-E_j^+\bigr) \,\,=\,\,
\frac{g\,M_j^{}}{\sqrt2\,m_W^{}}\bigl(U_{\nu N}^{}\bigr)_{ij}\,\bar u_l^{}P_{\rm R}^{}v_E^{} ~,
\end{eqnarray}
and so we arrive at
\begin{eqnarray} \label{w_h2lE}
\Gamma(h\to l^-E^+) \,\,=\,\, \sum_{i,j=1}^3 \Gamma\bigl(h\to l_i^-E_j^+\bigr)
\,\,=\,\,
\sum_i \frac{g^2m_h^{}M_i^2\,\bigl(U_{\nu N}^\dagger U_{\nu N}^{}\bigr)_{ii}}{32\pi\,m_W^2}
\Biggl(1-\frac{M_i^2}{m_h^2}\Biggr)^{\!\!2} ~,
\end{eqnarray}
having used the fact that $N$ and $E$ in each triplet have the same mass and
neglected the mass of~$l$.  Similarly, \,$\Gamma(h\to l^+E^-)=\Gamma(h\to l^-E^+)$.\,
Comparing Eqs.~(\ref{w_h2nN}) and~(\ref{w_h2lE}), we see that
\,$\Gamma(h\to\nu N)=\Gamma(h\to l^-E^+)$.\,

As in the type-I case, there are experimental constraints that the elements of $U_{\nu N}$ must
satisfy, besides the requirement in Eq.~(\ref{UmU}).
Expressed in terms of \,$\epsilon=U_{\nu N}^{}U_{\nu N}^\dagger$\, as before, in type-III
seesaw the bounds extracted from electroweak precision data are~\cite{delAguila:2008cj,ewpd}
\begin{eqnarray} \label{eii'}
\epsilon_{11}^{} \,\,\le\,\, 3.6\times10^{-4} ~, \hspace{5ex}
\epsilon_{22}^{} \,\,\le\,\, 2.9\times10^{-4} ~, \hspace{5ex}
\epsilon_{33}^{} \,\,\le\,\, 7.3\times10^{-4} ~,
\end{eqnarray}
whereas from the measurements of lepton-flavor violating transitions~\cite{Abada:2008ea}
\begin{eqnarray} \label{fcnc'}
|\epsilon_{12}^{}| \,\,\le\,\, 1.7\times10^{-7} ~, \hspace{5ex}
|\epsilon_{13}^{}| \,\,\le\,\, 4.2\times 10^{-4} ~, \hspace{5ex}
|\epsilon_{23}^{}| \,\,\le\,\, 4.9\times 10^{-4} ~.
\end{eqnarray}
In addition, direct searches for heavy charged leptons at colliders impose constraints on
the mass of $E$, and hence the mass of $N$ as well, namely
\,$M_i^{}\,\raisebox{-0.4ex}{\footnotesize$\stackrel{\textstyle>}{\sim}$}\,100$\,GeV\,~\cite{pdg}.

\begin{figure}[t]
\includegraphics[width=168mm]{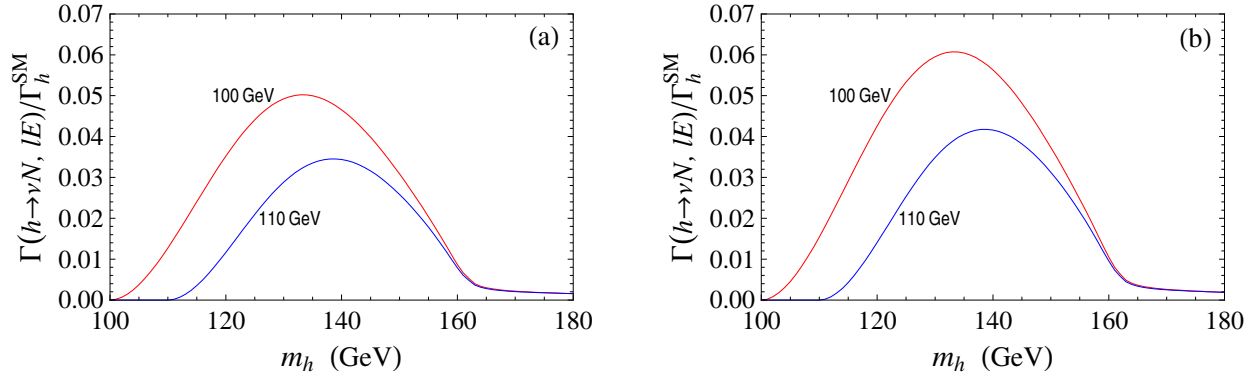} \vspace*{-2ex}
\caption{Ratios of the sum of \,$h\to\nu N$\, and \,$h\to l^\pm E^\mp$\, rates in type-III seesaw
to the total Higgs width in the SM as functions of the Higgs mass $m_h^{}$ for heavy-lepton
mass values  \,$m_N^{}=m_E^{}=100,110$~GeV\, and two different choices of $U_{\nu N}^{}$
as described in the text.\label{ratio3}} \vspace*{1ex}
\end{figure}

To explore the effect of large light-heavy mixing on the decays \,$h\to\nu N,lE$,\, we again
adopt some of the examples of $U_{\nu N}$ from Ref.~\cite{He:2009ua} for illustrations.
In addition, we assume that the three triplets are all degenerate,
\,$M_1^{}=M_2^{}=M_3^{}=m_N^{}=m_E^{}$.\,
As discussed in Ref.~\cite{He:2009ua}, the choices of $U_{\nu N}$ in Eqs.~(\ref{u0e})
and~(\ref{u0f}) are also appropriate for type-III seesaw, as they yield \,$\epsilon_{12}^{}=0$,\,
automatically fulfilling the very stringent requirement on $\epsilon_{12}^{}$ in Eq.~(\ref{fcnc'}).
For the first one with \,$b=a$,\, we obtain its largest allowed value to be \,$a=0.012$.\,
This leads to the plot in Fig.~\ref{ratio3}(a) which shows the ratio of the rate sum
\,$\Gamma(h\to\nu N)+\Gamma(h\to l^+E^-)+\Gamma(h\to l^-E^+)$\, to the Higgs total width
$\Gamma_h^{\rm SM}$ in the SM as functions of the Higgs mass $m_h^{}$ for
\,$m_N^{}=100,110$~GeV.\,
For the choice of $U_{\nu N}$ as in Eq.~(\ref{u0f}) with \,$b=(0.0013+1.03\,i)a$,\, we find
that the maximum allowed value \,$a=0.013$\, results in somewhat greater rates, as can be
seen in Fig.~\ref{ratio3}(b).

Thus in type-III seesaw the large light-heavy mixing gives rise to modifications of the SM Higgs
total width that are modest, roughly only~5\%, much smaller than those in the type-I case.
This is due to the stronger experimental constraints on the elements of the mixing matrix
$U_{\nu N}$ and also to the lower-limit on the heavy-lepton masses.
As a consequence, the Higgs decays into four light fermions in this case would be less
sensitive for probing the underlying seesaw mechanism than their type-I counterparts.

\section{Conclusions}

In seesaw scenarios with more than one generation of light and heavy neutrinos, it is possible
in some circumstances to have sizable mixing between them for heavy-neutrino masses of order
100\,GeV or less.
We have explored this possibility further, taking into account constraints from currently
available experimental data, and considered its effect on Higgs-boson decays
in the contexts of the seesaw models of types~I and~III.

We have found that the Higgs decay into a pair of light and heavy neutrinos, \,$h\to\nu N$,\,
in type-I seesaw with large light-heavy mixing could enhance the total Higgs width in
the standard model by up to nearly 30\% for a relatively light Higgs-boson,
with \,$m_h^{}\mbox{\footnotesize\,$\lesssim$\,}140$\,GeV.\,
This would imply sizable reduction in branching ratio for some of the important Higgs decay
modes in the SM and thus could significantly affect Higgs searches at the LHC or other colliders.
We have shown that the $N$ produced in \,$h\to\nu N$\, very quickly decays into three
light fermions, which makes this Higgs decay effectively a four-body decay.
We have further found that, in type-I seesaw with the large mixing, these four-body Higgs
decays can have rates comparable to, or a few times larger than, their SM counterparts and
therefore may provide a potentially important avenue to uncover the underlying seesaw
mechanism.
In type-III seesaw, because of stricter experimental constraints the corresponding
\,$h\to\nu N$\, decays and the decays involving the charged leptons, \,$h\to l^\pm E^\mp$,\,
produce only modest, of order~5\%, enlargement of the total Higgs width in the~SM.
All these considerations suggest that, if the Higgs boson is discovered and its decays can be
well measured, its decays into a~light neutrino plus three other light fermions could
serve to probe the seesaw models.
Hence such Higgs decays could yield information complementary to that possibly available from
the direct searches for these heavy leptons at colliders.

\acknowledgments

This work was supported in part by NSC and NCTS.
X.G.H. thanks IPMU for hospitality while part of this work was carried out.

\appendix

\section{Amplitudes for decays of \boldmath$N$ into three light fermions}

The decays of $N$ for \,$m_N^{}<m_W^{}$\, are mainly into three light fermions.
If  \,$m_h^{}>m_N^{}>m_W^{}$ or~$m_Z^{}$,\, the main $N$ decays are effectively still
three-body, as the daughter $W$ or $Z$ will quickly decay into a pair of light fermions.
Thus the relevant diagrams are each mediated by the $W$, $Z$, or Higgs boson.
In calculating their decay rates we neglect the light-fermion masses, and so in the amplitudes
below we drop terms proportional to light-fermion masses.
Consequently, since the Higgs coupling to a light fermion is proportional to its mass,
we also drop the contributions of the Higgs-mediated diagrams.

From ${\cal L}'$ in Eq.~(\ref{L'}), we obtain the amplitude for
\,$N_i^{}\to\nu_j^{}(p_\nu')\,\nu_k^{}(p_\nu^{})\,\bar\nu_k^{}(p_{\bar\nu}^{})$
\begin{eqnarray}
{\cal M}\bigl(N_i^{}\to\nu_j^{}\nu_k^{}\bar\nu_k^{}\bigr) &=&
\frac{g^2}{4c_{\rm w}^2}\, \frac{\bar u_\nu'\gamma^\alpha\Bigl[
\bigl(U_{\nu\nu}^\dagger U_{\nu N}^{}\bigr)_{ji}P_{\rm L}^{} -
\bigl(U_{\nu\nu}^{\rm T}U_{\nu N}^*\bigr)_{ji}P_{\rm R}^{} \Bigr] u_N^{}\,
\bar u_\nu^{}\gamma_\alpha^{}P_{\rm L}^{}v_\nu^{}}
{m_Z^2-\bigl(p_\nu^{}+p_{\bar\nu}^{}\bigr)^2-i\Gamma_Z^{}m_Z^{}} ~,
\end{eqnarray}
where no summation over $k$ is implied and we have
employed~\,$U_{\nu\nu}^\dagger U_{\nu\nu}^{}\simeq1$.\,
In deriving this and the other amplitudes below, we make use of the Majorana nature of
the neutrinos, \,$\nu=\nu^c$\, and~\,$N=N^c$.\,
Thus, for \,$N_i^{}\to\nu_j^{}(p_\nu^{})\,l_m^-(p_-^{})\,l_n^+(p_+^{})$\, with
\,$m\neq n$,\, we find
\begin{eqnarray}
{\cal M}\bigl(N_i^{}\to\nu_j^{}l_m^-l_n^+\bigr) &=&
\frac{g^2}{2}\, \frac{\bigl(U_{\nu N}^{}\bigr)_{m i}\bigl(U_{\nu\nu}^\dagger\bigr)_{j n}\,\,
\bar u_l^{}\gamma^\alpha P_{\rm L}^{}u_N^{}\,\bar u_\nu^{}\gamma_\alpha^{}P_{\rm L}^{}v_l^{}}
{m_W^2-\bigl(p_\nu^{}+p_+^{}\bigr)^2-i\Gamma_W^{}m_W^{}}
\nonumber \\ && \!\! +\;
\frac{g^2}{2}\, \frac{\bigl(U_{\nu N}^*\bigr)_{n i}\bigl(U_{\nu\nu}^{\rm T}\bigr)_{j m}\,\,
\bar u_\nu^{}\gamma^\alpha P_{\rm R}^{}u_N^{}\,\bar u_l^{}\gamma_\alpha^{}P_{\rm L}^{}v_l^{}}
{m_W^2-\bigl(p_\nu^{}+p_-^{}\bigr)^2-i\Gamma_W^{}m_W^{}} ~,
\end{eqnarray}
where in the second term we have performed a Fierz transformation and a matrix
transposition of the charged-lepton part.
For \,$N_i^{}\to\nu_j^{}(p_\nu^{})\,l_k^-(p_-^{})\,l_k^+(p_+^{})$,\,  we have
\begin{eqnarray} \label{N2nll}
{\cal M}\bigl(N_i^{}\to\nu_j^{}l_k^-l_k^+\bigr) &=&
\frac{g^2}{2}\, \frac{\bigl(U_{\nu N}^{}\bigr)_{k i}\bigl(U_{\nu\nu}^\dagger\bigr)_{j k}\,\,
\bar u_l^{}\gamma^\alpha P_{\rm L}^{}u_N^{}\,\bar u_\nu^{}\gamma_\alpha^{}P_{\rm L}^{}v_l^{}}
{m_W^2-\bigl(p_\nu^{}+p_+^{}\bigr)^2-i\Gamma_W^{}m_W^{}}
\nonumber \\ && \!\!\! +\;
\frac{g^2}{2}\, \frac{\bigl(U_{\nu N}^*\bigr)_{k i}\bigl(U_{\nu\nu}^{\rm T}\bigr)_{j k}\,\,
\bar u_\nu^{}\gamma^\alpha P_{\rm R}^{}u_N^{}\,\bar u_l^{}\gamma_\alpha^{}P_{\rm L}^{}v_l^{}}
{m_W^2-\bigl(p_\nu^{}+p_-^{}\bigr)^2-i\Gamma_W^{}m_W^{}}
\nonumber \\ && \!\!\! -\;
\frac{g^2}{2c_{\rm w}^2}\,\frac{\bar u_\nu^{}\gamma^\alpha \Bigl[
\bigl(U_{\nu\nu}^\dagger U_{\nu N}^{}\bigr)_{ji} P_{\rm L}^{} -
\bigl(U_{\nu\nu}^{\rm T} U_{\nu N}^*\bigr)_{ji} P_{\rm R}^{} \Bigr] u_N^{}\,
\bar u_l^{}\gamma_\alpha^{}\bigl(l_l^{}P_{\rm L}^{}+r_l^{}P_{\rm R}^{}\bigr)v_l^{}}
{m_Z^2-\bigl(p_+^{}+p_-^{}\bigr)^2-i\Gamma_Z^{}m_Z^{}} ~, ~~~~~~~
\end{eqnarray}
where no summation over $k$ is implied, \,$l_l^{}=-\frac{1}{2}+s_{\rm w}^2$,\, and
\,$r_l^{}=s_{\rm w}^2$,\, with \,$s_{\rm w}^{}=\sin\theta_{\rm W}$.\,

There are also decays into final states containing a lepton and a pair of quarks.
We derive for
\,$N_i^{}\to\nu_j^{}(p_\nu^{})\,q\bigl(p_q^{}\bigr)\,\bar q\bigl(p_{\bar q}^{}\bigr)\,$
\begin{eqnarray}
{\cal M}\bigl(N_i^{}\to\nu_j^{}q\bar q\bigr) &=&
\frac{g^2}{2c_{\rm w}^2}\, \frac{\bar u_\nu^{}\gamma^\alpha \Bigl[
\bigl(U_{\nu\nu}^\dagger U_{\nu N}^{}\bigr)_{ji}P_{\rm L}^{} -
\bigl(U_{\nu\nu}^{\rm T}U_{\nu N}^*\bigr)_{ji}P_{\rm R}^{} \Bigr]u_N^{}\,
\bar u_q^{}\gamma_\alpha^{}\bigl(l_q^{}P_{\rm L}^{}+r_q^{}P_{\rm R}^{}\bigr)v_q^{}}
{m_Z^2-\bigl(p_q^{}+p_{\bar q}^{}\bigr)^2-i\Gamma_Z^{}m_Z^{}} ~, ~~~~
\end{eqnarray}
where $q$ can be an up-type quark $u$ or down-type quark $d$, with
\begin{eqnarray}
l_u^{} \;=\; \mbox{$\frac{1}{2}$}-\mbox{$\frac{2}{3}$}\,s_{\rm w}^2 ~, ~~~~~
r_u^{} \;=\;-\mbox{$\frac{2}{3}$}\,s_{\rm w}^2 ~, \hspace{5ex}
l_d^{} \;=\; -\mbox{$\frac{1}{2}$}+\mbox{$\frac{1}{3}$}\,s_{\rm w}^2 ~, ~~~~~
r_d^{} \;=\; \mbox{$\frac{1}{3}$}\,s_{\rm w}^2 ~.
\end{eqnarray}
For \,$N_i^{}\to l_j^-(p_-^{})\,u(p_u^{})\bar d\bigl(p_d^{}\bigr)$\,  we find
\begin{eqnarray}
{\cal M}\bigl(N_i^{}\to l_j^-u\bar d\bigr) &=&
\frac{g^2}{2}\, \frac{\bigl(U_{\nu N}^{}\bigr)_{ji}V_{ud}^{}\,\,
\bar u_l^{}\gamma^\alpha P_{\rm L}^{}u_N^{}\,\bar u_u^{}\gamma_\alpha^{}P_{\rm L}^{}v_d^{}}
{m_W^2-\bigl(p_u^{}+p_d^{}\bigr)^2-i\Gamma_W^{}m_W^{}} ~,
\end{eqnarray}
where $V_{ud}$ is an element of the Cabibbo-Kobayashi-Maskawa (CKM) matrix.
The amplitude for \,$N_i^{}\to l_j^+\bar u d$\, is similar in form.

\end{document}